\documentclass[nofootinbib,a4paper,aps,prd,10pt,superscriptaddress,showkeys,twocolumn]{revtex4}

\usepackage{graphicx}
\usepackage{amsfonts}
\usepackage{amssymb} 
\usepackage{amsmath}
\usepackage{hyperref}
\usepackage{natbib}
\usepackage{float}
\usepackage{dcolumn}

\usepackage{bm}% bold math
\usepackage{latexsym,color}

\newcommand{\bea}{\begin{eqnarray}}
\newcommand{\eea}{\end{eqnarray}}
\newcommand{\be}{\begin{equation}}
\newcommand{\ee}{\end{equation}}

\begin{document}

\preprint{APS/123-QED}

\title{Energy transfer between the sources in gravitational decoupling }

\author{Daulet Berkimbayev}
\email{daulet9432@gmail.com}
\affiliation{Al-Farabi Kazakh National University, 71 Al-Farabi Ave., Almaty 050040, Kazakhstan}
\affiliation{National Nanotechnology Laboratory of Open Type,  Almaty 050040, Kazakhstan.}

\begin{abstract}
A straightforward and fully analytic approach is introduced to examine how polytropic fluids influence arbitrary gravitational sources in static, spherically symmetric spacetimes. As a concrete application, we explore the internal mechanism of energy transfer between gravitational sources embedded within a self-gravitating system.
\end{abstract}


\maketitle

\section{Introduction}

The investigation of self-gravitating matter configurations continues to play a pivotal role in the theoretical framework of general relativity. In particular, the study of the internal dynamics of compact stellar objects is essential for a deeper comprehension of gravitational collapse mechanisms and their associated phenomenology. From a physical standpoint, it is reasonable to expect that the interior of such compact structures is composed of a conglomeration of various types of matter and field content, typically modeled as a multi-fluid system. These constituents may include, but are not limited to, anisotropic fluids, dissipative components, charged matter, or scalar fields, all of which may interact in complex and non-linear ways.

Despite the mathematical convenience of describing the total matter content via a single effective energy-momentum tensor, denoted by $\tilde{T}_{\mu\nu}$, as required by the Einstein field equations,
\begin{equation}
\tilde{T}_{\mu\nu} = T^{(1)}_{\mu\nu} + T^{(2)}_{\mu\nu} + \cdots + T^{(n)}_{\mu\nu},
\label{total_tmn}
\end{equation}
such a unification may obscure important features of the system, especially when the internal constituents differ substantially in their thermodynamic behavior or coupling to geometry. Equation \eqref{total_tmn} expresses the principle that gravity couples to the total energy-momentum distribution, irrespective of its decomposition into physically distinguishable sources.

However, the practice of aggregating all internal degrees of freedom into a single effective description is often guided not by physical necessity but by the aim of mathematical tractability. In doing so, one risks introducing oversimplified models that, while potentially yielding exact or analytically tractable solutions, may fall short of capturing the physical complexity inherent in realistic astrophysical environments. A common example is the imposition of a simplistic equation of state across the entire configuration, which may inadequately represent systems with localized anisotropies, charge distributions, or transport phenomena.

While such an approach can be justified in many idealized models—leading to viable and regular solutions—it must be recognized that this simplification process entails a significant reduction in the system’s dynamical degrees of freedom. This reduction may inadvertently suppress key physical processes, including energy transfer mechanisms between subsystems, pressure anisotropy evolution, or dissipative effects. As a result, the obtained model, though mathematically elegant, may offer an idealized or even misleading picture of the underlying physics.

It is thus imperative to develop frameworks that preserve the individuality of the constituent sources while ensuring compatibility with the gravitational field equations. This requirement motivates the use of methods such as gravitational decoupling, which allow the splitting of the total system into interacting sectors with distinct physical interpretations and mutual backreaction. Such an approach enables a more refined and physically consistent analysis of the structure and evolution of self-gravitating configurations.

Given the discussion above, it becomes evident that one of the crucial tasks in the modeling of self-gravitating systems is the identification and quantification of the individual contributions of each matter component to the overall structure and dynamics of the configuration. Understanding the gravitational behavior of each constituent source and how these components interact is particularly relevant for determining the dominant physical mechanism within the system, as well as for assessing the compatibility of imposed equations of state with the prevailing matter content.

If a particular fluid or field dominates the dynamics in a given region, then any equation of state imposed globally on the system should remain consistent with this dominant sector. Otherwise, the model may yield physically untenable or unstable configurations. However, due to the non-linear and coupled structure of Einstein’s field equations, disentangling the individual role of each source in a generic multi-fluid configuration is highly non-trivial. In conventional general relativity, the entanglement of sources within the total energy-momentum tensor often precludes such a decomposition without significant approximations or the application of numerical techniques.

To address this limitation, the framework of Gravitational Decoupling (GD), as developed in~\cite{Ovalle2017, Ovalle2019}, offers a novel and systematic method for isolating and analyzing the behavior of individual gravitational sources. The GD method enables the analytical decoupling of Einstein's equations into simpler sub-systems, each associated with one component of the total energy-momentum content. Crucially, this can be accomplished without resorting to perturbative expansions or numerical algorithms, making the approach particularly valuable for the study of exact solutions and analytic insights.

To illustrate the mechanism, let us specialize the general decomposition to the case of two interacting sources: a primary matter content $T_{\mu\nu}$ and an auxiliary source $\theta_{\mu\nu}$. The total energy-momentum tensor is then given by
\begin{equation}
\tilde{T}_{\mu\nu} = T_{\mu\nu} + \theta_{\mu\nu}.
\label{eq:2source}
\end{equation}
Since Einstein’s equations imply the vanishing divergence of the total energy-momentum tensor,
\begin{equation}
\nabla^{\mu} \tilde{T}_{\mu\nu} = \nabla^{\mu} T_{\mu\nu} + \nabla^{\mu} \theta_{\mu\nu} = 0,
\label{eq:bianchi_full}
\end{equation}
the interaction between the two sectors is governed by the divergence properties of the individual components. There are two distinct possibilities consistent with \eqref{eq:bianchi_full}:
\begin{align}
\text{(i)} \quad & \nabla^{\mu} T_{\mu\nu} = 0, \quad \nabla^{\mu} \theta_{\mu\nu} = 0, \label{eq:conserved_case}\\
\text{(ii)} \quad & \nabla^{\mu} T_{\mu\nu} = -\nabla^{\mu} \theta_{\mu\nu} \neq 0. \label{eq:interaction_case}
\end{align}

In the first scenario~\eqref{eq:conserved_case}, each source is independently conserved, indicating that the interaction is exclusively gravitational, mediated by the spacetime curvature without direct energy or momentum exchange. This is often considered an idealized or decoupled limit, suitable for simplifying assumptions but potentially missing key dynamical effects.

The second scenario~\eqref{eq:interaction_case}, which is generally more realistic, entails a non-trivial exchange of energy and momentum between the two sectors. This represents a genuine interaction and implies the presence of coupling mechanisms beyond pure gravity—such as scalar field interactions, fluid mixing, or effective forces arising from internal structure. However, because the contracted Bianchi identities are identically satisfied by virtue of the geometric structure of general relativity, they do not by themselves provide any additional dynamical information or constraints on the nature of the interaction. That is, \eqref{eq:bianchi_full} is not an evolution equation, but rather a consistency condition inherited from the Einstein tensor's divergence-free nature.

The innovation of the Gravitational Decoupling method lies precisely in its ability to circumvent the triviality of the Bianchi identities by explicitly restructuring the field equations in such a way that the decoupled contributions from $T_{\mu\nu}$ and $\theta_{\mu\nu}$ can be solved sequentially or in parallel, depending on the context. This permits a controlled and tractable exploration of energy exchange phenomena, making it possible to derive effective interaction terms, interpret them physically, and constrain the resulting configurations analytically.

In the sections that follow, we will explore how this framework can be employed to examine energy transfer processes in a variety of self-gravitating systems, including anisotropic fluids and non-minimally coupled scalar fields, and how the dynamical role of each constituent source can be made explicit within the gravitational decoupling scheme.

Among the many possible configurations of matter content, we devote special attention in this work to a gravitational source governed by a polytropic equation of state. The decision to focus on this class stems not only from its historical relevance in astrophysical modeling but also from its broad applicability in describing diverse compact objects across different regimes of density and pressure. 

Our objective is to investigate how such a polytropic component, once coupled through gravitational interaction, can influence the behavior and structure of an accompanying fluid distribution, whose physical nature remains unspecified. This strategy enables a general and flexible assessment of interaction dynamics, without being constrained by the particular properties of the secondary source. By doing so, we aim to extract model-independent insights into how a well-characterized gravitational contributor such as a polytrope modifies the overall structure of a composite system.

We begin by recalling the main aspects of the Gravitational Decoupling (GD) method in the context of static, spherically symmetric spacetimes, as formulated in detail in~\cite{Ovalle2019}. For its generalization to axially symmetric geometries, the reader is referred to~\cite{Contreras2021}. The GD framework, and in particular its most basic realization based on the Minimal Geometric Deformation (MGD) approach~\cite{Ovalle2020,DaRocha2017a,DaRocha2017b,FernandesSilva2018a,Casadio2018,FernandesSilva2018b,Contreras2018a,Panotopoulos2018,DaRocha2019,LasHeras2019,Rincon2019,DaRocha2020,Contreras2020,Arias2020,DaRocha2020b,TelloOrtiz2020,DaRocha2020c,Meert2021,TelloOrtiz2021,Maurya2021,Azmat2021,Maurya2022,Ovalle2018b}, has attracted considerable interest in the last decade. A wide range of studies have applied it in diverse physical settings, both within and beyond general relativity~\cite{Heras2018,Gabbanelli2018,Azmat2021b,Azmat2021a,Maurya2022b,Morales2018,Sharif2018a,Ovalle2018,Estrada2018,Estrada2019,Sharif2018b,Gabbanelli2019,Ovalle2019b,Ovalle2019b,Linares2020,Sharif2018c,Ovalle2018b,Estrada2019b,Hensh2019,TelloOrtiz2020b,Sharif2020b,Sharif2020,Singh2019,Casadio2019b,Maurya2020b,Sharif2019,Maurya2019,Maurya2020,Ovalle2021b,Maurya2020c,Sharif2020c,Afrin2021,Islam2021,Zubair2020b,Maurya2021c,AmaTulMughani2021,Estrada2020b,Ovalle2021,Omwoyo2021,Sharif2021b,Maurya2022c,Afrin2022,Meert2022,Maurya2021d,DaRocha2022,DaRocha2021b,Sultana2021}.

Several compelling features make GD a versatile and powerful tool in gravitational modeling. The ability to couple multiple gravitational sources, enabling extensions of known solutions to include additional physical effects. The decomposition of a complex energy-momentum configuration into simpler, tractable components through systematic decoupling. Its applicability in modified gravity theories, allowing the derivation of novel solutions that lie beyond Einstein’s original framework.The generation of rotating and hairy black hole solutions, among other significant applications.

To set the stage, let us consider the Einstein field equations expressed as
\begin{equation}
G_{\mu\nu} \equiv R_{\mu\nu} - \frac{1}{2} R\, g_{\mu\nu} = \kappa\, \tilde{T}_{\mu\nu},
\label{eq:einstein_eqs}
\end{equation}
where the total energy-momentum tensor $\tilde{T}_{\mu\nu}$ is decomposed as
\begin{equation}
\tilde{T}_{\mu\nu} = T_{\mu\nu} + \theta_{\mu\nu}.
\label{eq:total_tensor}
\end{equation}
In this expression, $T_{\mu\nu}$ typically corresponds to a matter distribution associated with a known exact solution of Einstein’s equations, while $\theta_{\mu\nu}$ represents an additional source—possibly arising from new fields, anisotropies, or alternative gravitational sectors that are not described by classical general relativity.

Throughout this work we adopt geometric units in which $c = 1$ and $\kappa = 8\pi G_{\mathrm{N}}$, where $G_{\mathrm{N}}$ denotes Newton’s gravitational constant.

As a direct consequence of the Bianchi identities, the total energy-momentum tensor that sources the gravitational field must satisfy local conservation. This yields the covariant condition
\begin{equation}
\nabla_{\mu} \tilde{T}^{\mu\nu} = 0,
\label{eq:bianchi_conservation}
\end{equation}
ensuring consistency with the underlying diffeomorphism invariance of the spacetime manifold.

For static, spherically symmetric configurations, the geometry of the spacetime can be described by a line element of the form
\begin{equation}
ds^{2} = e^{\nu(r)} dt^{2} - e^{\lambda(r)} dr^{2} - r^{2}(d\theta^{2} + \sin^{2}\theta\, d\phi^{2}),
\label{eq:metric}
\end{equation}
where the metric functions $\nu(r)$ and $\lambda(r)$ depend solely on the areal radius $r$. The term $d\Omega^{2} = d\theta^{2} + \sin^{2}\theta\, d\phi^{2}$ represents the metric on the unit 2-sphere.

Substituting the metric \eqref{eq:metric} into the Einstein field equations~\eqref{eq:einstein_eqs}, and incorporating the decomposition of the total energy-momentum tensor as in~\eqref{eq:total_tensor}, one obtains the following system of field equations:
\begin{align}
\kappa (T_{00} + \theta_{00}) &= \frac{1}{r^{2}} - e^{-\lambda} \left( \frac{1}{r^{2}} - \frac{\lambda'}{r} \right), \label{eq:einstein_tt} \\
\kappa (T_{11} + \theta_{11}) &= \frac{1}{r^{2}} - e^{-\lambda} \left( \frac{1}{r^{2}} + \frac{\nu'}{r} \right), \label{eq:einstein_rr} \\
\kappa (T_{22} + \theta_{22}) &= -\frac{e^{-\lambda}}{4} \left( 2\nu'' + \nu'^{2} - \lambda'\nu' + \frac{2(\nu' - \lambda')}{r} \right). \label{eq:einstein_ttang}
\end{align}
Here, primes denote derivatives with respect to $r$.

By comparing the terms in Eqs.~\eqref{eq:einstein_tt}–\eqref{eq:einstein_ttang}, we can define the effective matter content of the spacetime as follows:
\begin{align}
\tilde{\rho} &= T_{00} + \theta_{00} = \rho + E, \label{eq:eff_density} \\
\tilde{p}_{r} &= -T_{11} - \theta_{11} = p_{r} + \mathcal{P}_{r}, \label{eq:eff_pr} \\
\tilde{p}_{t} &= -T_{22} - \theta_{22} = p_{t} + \mathcal{P}_{t}, \label{eq:eff_pt}
\end{align}
where we interpret $\rho$, $p_{r}$, and $p_{t}$ as the energy density, radial pressure, and tangential pressure of the primary matter source $T_{\mu\nu}$, and $E$, $\mathcal{P}_{r}$, and $\mathcal{P}_{t}$ as the corresponding contributions from the auxiliary source $\theta_{\mu\nu}$.

The full decomposition of the two energy-momentum tensors may thus be written as
\begin{align}
T_{\mu\nu} &= \text{diag}[\rho, -p_{r}, -p_{t}, -p_{t}], \label{eq:T_tensor}\\
\theta_{\mu\nu} &= \text{diag}[E, -\mathcal{P}_{r}, -\mathcal{P}_{t}, -\mathcal{P}_{t}]. \label{eq:theta_tensor}
\end{align}

In general, the effective system exhibits anisotropic pressure, characterized by a nonzero anisotropy function
\begin{equation}
\Pi \equiv \tilde{p}_{t} - \tilde{p}_{r}, \label{eq:anisotropy}
\end{equation}
which reflects the difference between tangential and radial stresses. As such, the gravitational system governed by Eqs.~\eqref{eq:einstein_tt}–\eqref{eq:einstein_ttang} must be treated as an anisotropic fluid distribution, even if the individual components are isotropic on their own.

To proceed with the gravitational decoupling procedure, we first consider a baseline configuration governed solely by the energy-momentum tensor $T_{\mu\nu}$, with the secondary source $\theta_{\mu\nu}$ absent. This corresponds to a spacetime in which the total energy-momentum tensor reduces to
\begin{equation}
\tilde{T}_{\mu\nu} = T_{\mu\nu} + \theta_{\mu\nu} \equiv T_{\mu\nu}, \quad \text{when} \quad \theta_{\mu\nu} = 0.
\label{eq:seed_source}
\end{equation}
The metric associated with this seed solution is assumed to have the standard static, spherically symmetric form:
\begin{equation}
ds^2 = e^{\xi(r)} dt^2 - e^{\mu(r)} dr^2 - r^2(d\theta^2 + \sin^2\theta\, d\phi^2),
\label{eq:seed_metric}
\end{equation}
where $\xi(r)$ and $\mu(r)$ are the temporal and radial metric functions, respectively.

The function $e^{-\mu(r)}$ encodes the gravitational potential and can be written in terms of the Misner–Sharp mass function $m(r)$ as
\begin{equation}
e^{-\mu(r)} = 1 - \frac{2m(r)}{r}, \qquad m(r) = \kappa \int_0^r x^2 T_{00}(x)\, dx.
\label{eq:mass_function}
\end{equation}

When the additional gravitational source $\theta_{\mu\nu}$ is introduced, its effect manifests as deformations in the geometry of the seed metric~\eqref{eq:seed_metric}. These deformations can be parametrized by two functions: $f(r)$ for the radial sector and $g(r)$ for the temporal sector. The full metric then takes the form
\begin{align}
\xi(r) &\longrightarrow \nu(r) = \xi(r) + g(r), \label{eq:temporal_deform}\\
e^{-\mu(r)} &\longrightarrow e^{-\lambda(r)} = e^{-\mu(r)} + f(r). \label{eq:radial_deform}
\end{align}
It is important to emphasize that these substitutions do not correspond to coordinate transformations. They represent genuine physical changes in the geometry induced by the new energy-momentum content.

The gravitational field equations can now be split into two distinct subsystems. The first, corresponding to the undeformed configuration governed by $T_{\mu\nu}$, reads
\begin{align}
\kappa \rho &= \frac{1}{r^2} - e^{-\mu} \left( \frac{1}{r^2} - \frac{\mu'}{r} \right), \label{eq:rho_T}\\
\kappa p_r &= -\frac{1}{r^2} + e^{-\mu} \left( \frac{1}{r^2} + \frac{\xi'}{r} \right), \label{eq:pr_T}\\
\kappa p_t &= \frac{e^{-\mu}}{4} \left( 2\xi'' + \xi'^2 - \mu'\xi' + \frac{2(\xi' - \mu')}{r} \right), \label{eq:pt_T}
\end{align}
which define the matter content of the seed solution and are assumed to be satisfied by the metric \eqref{eq:seed_metric}.

The second set of equations describes the contribution from the additional source $\theta_{\mu\nu}$, appearing as corrections due to the functions $f(r)$ and $g(r)$:
\begin{align}
\kappa E &= -\frac{f}{r^2} - \frac{f'}{r}, \label{eq:E_theta}\\
\kappa \mathcal{P}_r - \mathcal{Z}_1 &= f \left( \frac{1}{r^2} + \frac{\nu'}{r} \right), \label{eq:Pr_theta}\\
\kappa \mathcal{P}_t - \mathcal{Z}_2 &= \frac{f}{4} \left( 2\nu'' + \nu'^2 + \frac{2\nu'}{r} \right) + \frac{f'}{4} \left( \nu' + \frac{2}{r} \right), \label{eq:Pt_theta}
\end{align}
where the quantities $\mathcal{Z}_1$ and $\mathcal{Z}_2$ capture the coupling between $f$ and $g$:
\begin{align}
\mathcal{Z}_1 &= e^{-\mu} \frac{g'}{r}, \label{eq:Z1}\\
\mathcal{Z}_2 &= e^{-\mu} \left( 2g'' + g'^2 + \frac{2g'}{r} + 2\xi' g' - \mu' g' \right). \label{eq:Z2}
\end{align}
When $f = g = 0$, the deformation vanishes, and the second source $\theta_{\mu\nu}$ has no effect—recovering the seed solution exactly.

Of particular interest is the case where $g(r) = 0$, meaning only the radial component of the metric is altered. This scenario corresponds to the so-called "Minimal Geometric Deformation" (MGD) limit, as introduced in~\cite{Ovalle2017}, in which the backreaction of $\theta_{\mu\nu}$ is entirely absorbed by the radial deformation $f(r)$. The resulting field equations become significantly simplified, and $f(r)$ can be determined directly from $\theta_{\mu\nu}$ and the background solution.

It is worth noting that Eqs.~\eqref{eq:E_theta}–\eqref{eq:Pt_theta} still involve the seed functions $\xi(r)$ and $\mu(r)$, indicating that the full system remains coupled, albeit now in a structured and tractable manner. This coupling is expected, as both $T_{\mu\nu}$ and $\theta_{\mu\nu}$ ultimately describe interrelated contributions to the same spacetime geometry.

To complete the system, we turn to the conservation condition~\eqref{eq:bianchi_conservation}. Evaluated with the full deformed metric, it reads:
\begin{align}
&\left[ T_{11}' - \frac{\xi'}{2} (T_{00} - T_{11}) - \frac{2}{r} (T_{22} - T_{11}) \right] \nonumber  - \frac{g'}{2} (T_{00} - T_{11}) \\
&\quad +\left[ \theta_{11}' - \frac{\nu'}{2} (\theta_{00} - \theta_{11}) - \frac{2}{r} (\theta_{22} - \theta_{11}) \right] = 0.
\label{eq:conservation_full}
\end{align}
The first square bracket represents the divergence of $T_{\mu\nu}$ computed with respect to the covariant derivative defined by the seed metric. Importantly, this term is a linear combination of Eqs.~\eqref{eq:rho_T}–\eqref{eq:pt_T}, reinforcing the consistency of the framework.

As is well known, the Einstein tensor constructed from a given metric automatically satisfies the contracted Bianchi identities. Therefore, for the geometry defined by the seed metric \eqref{eq:seed_metric}, the corresponding Einstein tensor $G^{(\xi,\mu)}_{\mu\nu}$ guarantees the conservation of the seed energy-momentum tensor $T_{\mu\nu}$:
\begin{equation}
\nabla^{(\xi,\mu)}_{\sigma} T^{\sigma\nu} = 0.
\label{eq:seed_conservation}
\end{equation}

However, when we consider the full deformed metric incorporating the geometric shifts $g(r)$ and $f(r)$, the covariant derivative changes accordingly. In this new geometry, the divergence of $T_{\mu\nu}$ becomes
\begin{equation}
\nabla_{\sigma} T^{\sigma\nu} = \nabla^{(\xi,\mu)}_{\sigma} T^{\sigma\nu} - \frac{g'}{2} (T^0_0 - T^1_1)\, \delta^\nu_1,
\label{eq:T_divergence}
\end{equation}
where the second term captures the contribution due to the deformation of the temporal component of the metric.

Combining this with the total conservation law $\nabla_{\sigma} \tilde{T}^{\sigma\nu} = 0$, we find that the divergence of the additional source must balance the deformation-induced term:
\begin{equation}
\nabla_{\sigma} T^{\sigma\nu} = -\nabla_{\sigma} \theta^{\sigma\nu} = -\frac{g'}{2}(T^0_0 - T^1_1)\, \delta^\nu_1.
\label{eq:exchange}
\end{equation}

This result is particularly noteworthy: it reflects an explicit and exact energy exchange between the two sources, governed by the deformation function $g(r)$. Moreover, the structure of Eq.~\eqref{eq:exchange} mirrors the form of the field equations associated with $\theta_{\mu\nu}$, namely Eqs.~\eqref{eq:E_theta}–\eqref{eq:Pt_theta}. This confirms that the system permits an exact, non-perturbative decoupling of gravitational sources under the GD framework, without requiring any small parameter expansion in $f$ or $g$~\cite{Ovalle2020}.

\subsection*{Summary of the Decoupling Procedure}

To avoid ambiguity and reinforce the conceptual structure developed between Eqs.~\eqref{eq:einstein_eqs} and~\eqref{eq:exchange}, we summarize the decoupling process and its link to energy exchange as follows:

\begin{enumerate}
\item We begin with a spacetime sourced by a single energy-momentum tensor $T_{\mu\nu}$, governed by the Einstein field equations in the form of Eqs.~\eqref{eq:rho_T}–\eqref{eq:pt_T}, with metric functions $\{\xi(r), \mu(r)\}$.

\item A second gravitational source $\theta_{\mu\nu}$ is introduced, yielding the total energy-momentum tensor $\tilde{T}_{\mu\nu} = T_{\mu\nu} + \theta_{\mu\nu}$. This modifies the background geometry via the replacements $\xi \rightarrow \nu = \xi + g$, and $e^{-\mu} \rightarrow e^{-\lambda} = e^{-\mu} + f$, as given in Eqs.~\eqref{eq:temporal_deform} and~\eqref{eq:radial_deform}.

\item The resulting spacetime, now characterized by metric functions $\{\nu(r), \lambda(r)\}$, must satisfy the Einstein field equations for the total source $\tilde{T}_{\mu\nu}$. This requirement splits into two coupled subsystems:
\begin{itemize}
    \item Eqs.~\eqref{eq:rho_T}–\eqref{eq:pt_T} for $T_{\mu\nu}$ in the undeformed geometry,
    \item Eqs.~\eqref{eq:E_theta}–\eqref{eq:Pt_theta} for $\theta_{\mu\nu}$ in the deformed background.
\end{itemize}

\item The internal consistency of this entire construction hinges on the conservation law $\nabla_\mu \tilde{T}^{\mu\nu} = 0$, which leads directly to Eq.~\eqref{eq:exchange}. This equation captures the transfer of energy between $T_{\mu\nu}$ and $\theta_{\mu\nu}$ and ensures that their interaction remains compatible with general covariance.

\end{enumerate}

We highlight several key points that emerge from the gravitational decoupling method:
\begin{itemize}
    \item The GD scheme provides an exact, non-perturbative mechanism for separating gravitational sources, including anisotropic fluids and exotic matter sectors.
    \item Although $\theta_{\mu\nu}$ may represent additional fields or even modifications of gravity, the full analysis is performed strictly within the framework of general relativity.
    \item The two sets of equations—those for $T_{\mu\nu}$ and $\theta_{\mu\nu}$—are not independent but are interrelated through the deformation functions and the Bianchi identities.
    \item Successful decoupling is achieved if the energy exchange between the sectors satisfies the relation in Eq.~\eqref{eq:exchange}, ensuring overall consistency of the full gravitational system.
\end{itemize}

To ensure a physically consistent spacetime description, the interior and exterior solutions must be smoothly joined at the stellar boundary, located at $r = R$. The interior geometry, valid in the region $0 \leq r \leq R$, is given by the deformed metric introduced earlier:
\begin{equation}
ds^2 = e^{\nu^{-}(r)} dt^2 - \left(1 - \frac{2\tilde{m}(r)}{r} \right)^{-1} dr^2 - r^2 d\Omega^2,
\label{eq:interior_metric}
\end{equation}
where the effective interior mass function $\tilde{m}(r)$ includes the radial geometric deformation $f(r)$:
\begin{equation}
\tilde{m}(r) = m(r) - \frac{r}{2} f(r).
\label{eq:mass_modified}
\end{equation}
Here, $m(r)$ is the Misner–Sharp mass defined previously in Eq.~\eqref{eq:mass_function}.

The exterior spacetime, valid for $r > R$, is assumed to be the vacuum Schwarzschild solution:
\begin{equation}
ds^2 = \left(1 - \frac{2M}{r} \right) dt^2 - \left(1 - \frac{2M}{r} \right)^{-1} dr^2 - r^2 d\Omega^2,
\label{eq:schwarzschild_metric}
\end{equation}
where $M$ represents the total mass as measured by a distant observer.

To join the two regions across the hypersurface $\Sigma$ defined by $r = R$, we apply the Israel–Darmois matching conditions, which require continuity of both the first and second fundamental forms across $\Sigma$.

The metric components must be continuous at the matching radius, leading to the conditions:
\begin{align}
e^{\nu^{-}(R)} &= 1 - \frac{2M}{R}, \label{eq:match1} \\
e^{-\lambda^{-}(R)} &= 1 - \frac{2M}{R}. \label{eq:match2}
\end{align}

The stress-energy content must also satisfy the continuity of the second fundamental form, which requires that the radial pressure vanishes at the boundary when viewed from the exterior. This condition, expressed using the total energy-momentum tensor, takes the form:
\begin{equation}
\left[ \tilde{T}_{\mu\nu} r^{\nu} \right]_{\Sigma} = 0,
\label{eq:second_fundamental}
\end{equation}
where $r^{\mu}$ is the outward-pointing unit normal vector to the hypersurface $r = R$.

Using Einstein's equations, Eq.~\eqref{eq:second_fundamental} translates to the requirement that the total radial pressure at the boundary must vanish:
\begin{equation}
[p_r + \mathcal{P}_r]_{r=R} = 0.
\label{eq:pressure_condition}
\end{equation}
Introducing the shorthand notations $p_R \equiv p_r(R)$ and $P_R \equiv \mathcal{P}_r(R)$, this becomes:
\begin{equation}
p_R + P_R = 0.
\label{eq:pressure_vanish}
\end{equation}

Using the expression for the deformed radial pressure from the field equations, the total effective pressure at $r = R$ can be written as:
\begin{equation}
\tilde{p}_r(R) = p_R + \frac{f(R)}{\kappa} \left( \frac{1}{R^2} + \frac{\nu'(R)}{R} \right) + \frac{g'(R)}{\kappa R} e^{-\mu(R)}.
\label{eq:effective_pressure}
\end{equation}
Setting $\tilde{p}_r(R) = 0$ gives the final constraint at the stellar surface.

The three conditions
\begin{align}
e^{\nu^{-}(R)} &= 1 - \frac{2M}{R}, \\
e^{-\lambda^{-}(R)} &= 1 - \frac{2M}{R}, \\
\tilde{p}_r(R) &= 0,
\end{align}
are both necessary and sufficient to ensure that the interior metric governed by gravitational decoupling can be consistently matched to the Schwarzschild vacuum exterior.

\section{Energy transfer}

From Eq.~\eqref{eq:exchange}, we observe that a key feature of the coupling between the two matter sectors is the explicit expression for the exchange of energy and momentum. This interaction is captured by the deformation of the temporal metric component, and the corresponding energy transfer between the sources is given by
\begin{equation}
\Delta E = \frac{g'}{2} (\rho + p_r),
\label{eq:energy_transfer}
\end{equation}
where $\rho$ and $p_r$ denote the energy density and radial pressure of the seed fluid, respectively. This compact relation makes transparent the mechanism through which gravitational decoupling enables energy exchange, highlighting the central role played by the metric deformation function $g(r)$.
In order to deepen our understanding of how energy flows between coupled gravitational sources, we shall consider a range of possible expressions for the energy-momentum exchange term \(\Delta E\). Rather than fixing a specific form a priori, we aim to explore how different functional dependencies of \(\Delta E\) influence the dynamics and geometry of the system. This approach allows us to assess the range of viable interaction mechanisms that may occur between the seed energy-momentum tensor \(T_{\mu\nu}\) and the additional contribution \(\theta_{\mu\nu}\), each of which may correspond to distinct physical fluids, fields, or modifications of gravity. By varying the form of \(\Delta E\), we can systematically examine how energy transfer modifies the effective matter content, deforms the metric functions, and alters the overall structure of the self-gravitating configuration.

\begin{figure*}[ht]
\center{\includegraphics[width=0.23\linewidth]{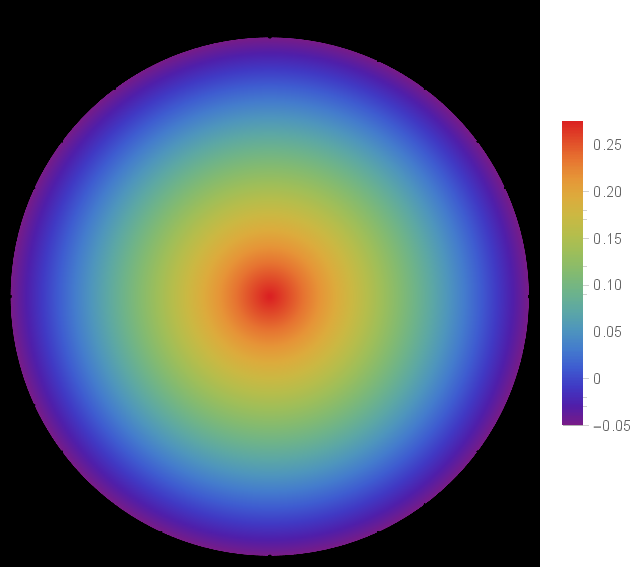}\quad \includegraphics[width=0.23\linewidth]{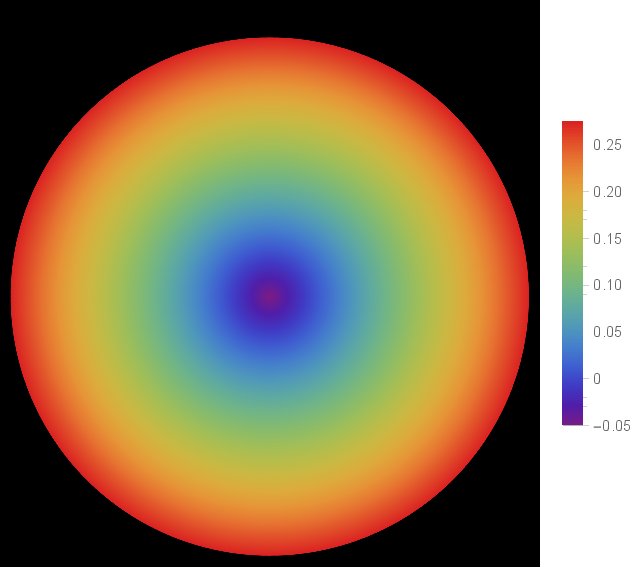}\quad
\includegraphics[width=0.23\linewidth]{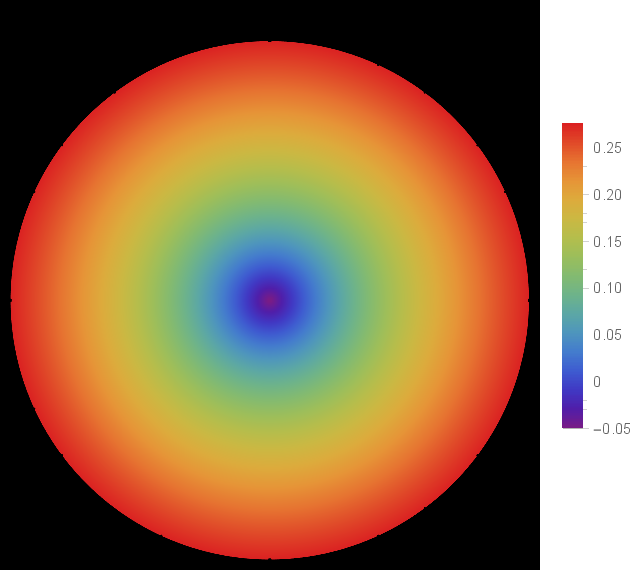}\quad 
\includegraphics[width=0.23\linewidth]{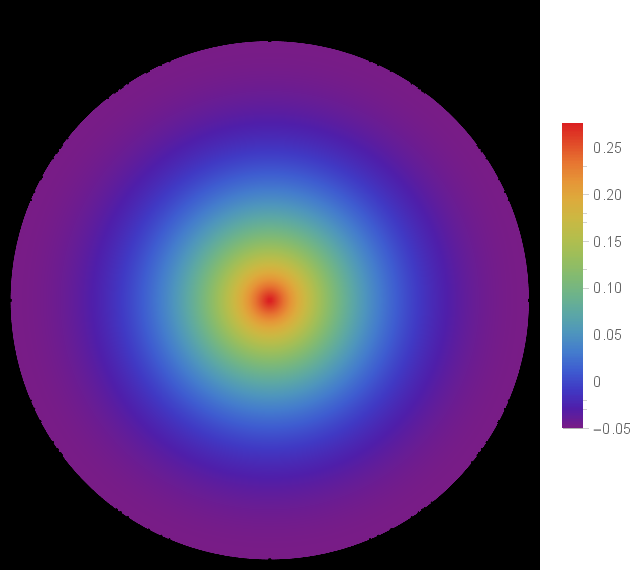}}\\
\center{\includegraphics[width=0.23\linewidth]{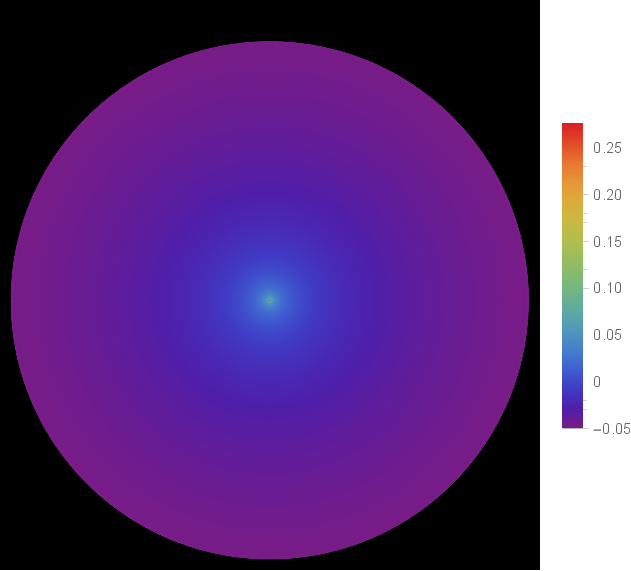}\quad \includegraphics[width=0.23\linewidth]{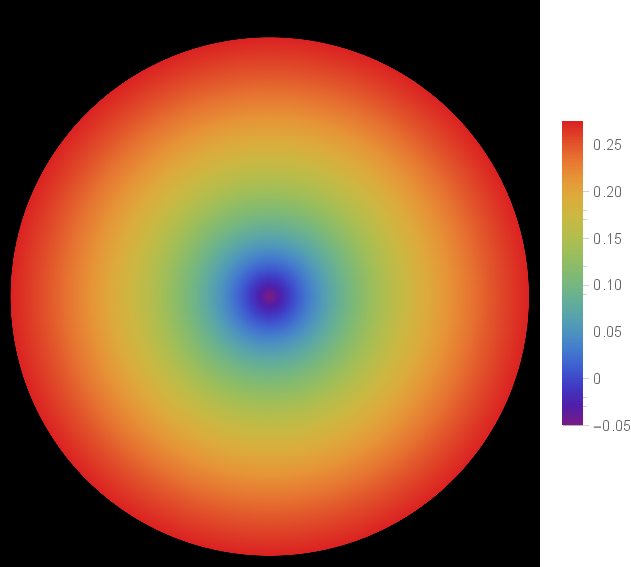}\quad \includegraphics[width=0.23\linewidth]{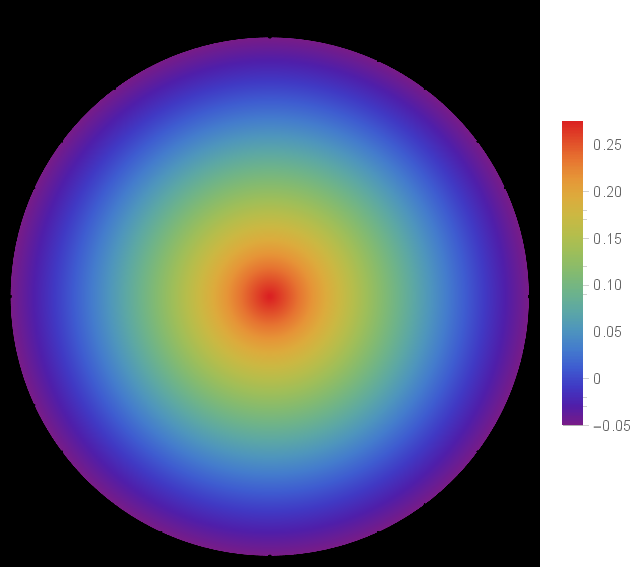}\quad
\includegraphics[width=0.23\linewidth]{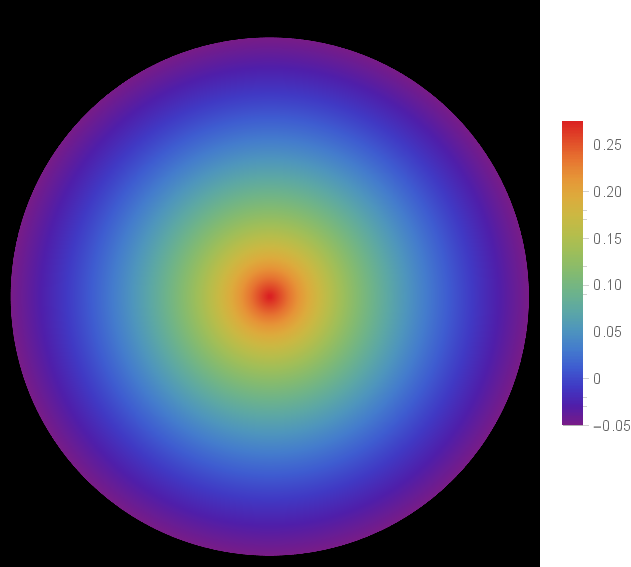}}
\caption{The spatial dependence of the
energy–exchange term \(\Delta E(r)\) for the eight suggested profiles.
The panels are ordered from left to right in the first row and then in
the second row. The top row displays, respectively, the
power-law growth profile \(\Delta E_1\),
the inverted power-law profile \(\Delta E_2\),
the exponential decay profile \(\Delta E_3\),
and the peaked profile \(\Delta E_4\).
The bottom row shows, in order, the logarithmic profile \(\Delta E_5\),
the smooth-decay profile \(\Delta E_6\),
the oscillatory profile \(\Delta E_7\),
and the saturating profile \(\Delta E_8\).}
\label{EnergyTransferColor}
\end{figure*}

In order to study the effect of energy transfer between decoupled gravitational sources, we introduce a family of spatially dependent exchange terms \(\Delta E(r)\), each representing a different hypothesis about how energy is redistributed within the interior of a compact self-gravitating object. In all cases the exchange term depends on the radial coordinate \(r\), the stellar radius \(R\), and two free parameters: a coupling strength \(C_E\) and an exponent \(n\). The candidate profiles are chosen to capture a wide variety of qualitatively distinct behaviors and to explore their physical plausibility in relativistic stellar configurations:

\begin{enumerate}
    \item \textbf{Power-law growth:} \(\displaystyle \Delta E_1 = C_E \left( \frac{r}{R} \right)^n\).  
    This profile increases monotonically from the center outward. For \(n>0\), it implies minimal interaction near the core and maximal energy exchange toward the surface. It models scenarios where the interaction between sources strengthens away from the core, for instance due to gradients in matter density, increasing anisotropy, or the dominance of one fluid in the outer layers. Such behavior can be particularly relevant for systems where surface layers play a significant dynamical role, such as strange stars or compact anisotropic objects.

    \item \textbf{Inverted power-law:} \(\displaystyle \Delta E_2 = C_E \left( 1 - \frac{r}{R} \right)^n\).  
    This decreasing function represents maximal energy exchange near the center, gradually diminishing toward the boundary. It is natural for modeling strong core-level coupling, possibly associated with dense nuclear matter, central phase transitions, or pressure-driven interactions between fluids at small radii. This form is often more realistic for stars with high central densities.

    \item \textbf{Exponential decay:} \(\displaystyle \Delta E_3 = C_E\, \exp\!\left( 1 - \frac{r}{R} \right)\).  
    An exponential falloff yields a smooth but rapid decay of energy exchange from the center outward. It shares qualitative features with \(\Delta E_2\), but provides sharper gradients, allowing finer control of spatial localization. This profile is well suited to systems where short-range forces or strongly localized coupling mechanisms dominate.

    \item \textbf{Peaked distribution:} \(\displaystyle \Delta E_4 = C_E\, r^n \exp\!\left( -\frac{r}{R} \right)\).  
    This expression rises from the origin, reaches a maximum at an intermediate radius, and decays toward the surface, forming a shell-like peak. Such a shape is physically compelling for systems in which energy exchange is localized neither in the core nor near the outer boundary, but rather in an intermediate interaction zone. It can model, for example, crust–core boundaries, mixed-phase layers, or thermally unstable mid-regions of a star.

    \item \textbf{Logarithmic form:} \(\displaystyle \Delta E_5 = C_E\, \ln\!\left( \frac{r}{R} \right)\).  
    The logarithmic profile is singular at the origin and negative for \(r < R\), so its direct physical application is limited unless some regularization is introduced. Nevertheless, it can be used as an idealized probe of highly localized core interactions, or as a test function for studying the stability of models with singular transfer terms. It may also serve as a toy model for systems near marginally stable phase transitions or collapse thresholds where central divergences play a role.

    \item \textbf{Smooth decay:} \(\displaystyle \Delta E_6 = \frac{C_E}{1 + \left( \frac{r}{R} \right)^n}\).  
    This form describes a gentle and bounded decrease in energy transfer from the core to the surface. It is regular everywhere and smoothly approaches small values at large \(r\), making it suitable for stable fluid–fluid interaction scenarios with mild gradients. Its smoothness and boundedness are particularly attractive for numerical implementations and stability analyses.

    \item \textbf{Oscillatory profile:} \(\displaystyle \Delta E_7 = C_E \sin\!\left( \frac{r}{R} \right)\).  
    Here the energy exchange alternates in sign, introducing successive zones of positive and negative transfer. This behavior could mimic resonant or wave-like coupling between gravitational sources, and may be applicable in exotic matter models, dynamically oscillating interiors, or configurations coupled to scalar fields supporting standing-wave solutions. Although mathematically regular over \(r \in [0,R]\), it typically requires a specific physical mechanism to justify its oscillatory nature.

    \item \textbf{Saturating profile:} \(\displaystyle \Delta E_8 = \frac{C_E \left( \frac{r}{R} \right)^n}{1 + \left( \frac{r}{R} \right)^n}\).  
    This sigmoidal profile rises from negligible coupling at the core and asymptotically approaches a constant value \(C_E\) toward the surface. It captures the intuition of an interaction that “switches on’’ once a certain density or pressure threshold is reached, making it suitable for effective models with activation-like energy transfer mechanisms or plateau-like interaction layers.
\end{enumerate}

Among these profiles, the most physically appealing in stellar applications are those that remain regular at the center (\(r=0\)) and allow for a smooth matching to the exterior solution at the surface (\(r=R\)), without introducing unbounded or strongly pathological behavior in the interior. This includes, in particular, the decreasing or localized profiles (2), (3), (4), (6), and (8), which naturally accommodate finite observables throughout the star and can be tuned to vanish or become small at the boundary. By contrast, the logarithmic form (5) is singular at the origin and thus mainly useful as a diagnostic or toy model, while the sinusoidal profile (7), though regular, typically requires additional physical input to motivate its oscillatory pattern.

Profiles with peaked or plateau-like structure, such as (4) and (8), are especially well suited to describing localized energy-exchange zones, for instance near phase transitions or in interaction layers separating distinct fluid regions. Each of the above functional forms is therefore intended to highlight a different qualitative aspect of gravitational interaction in a decoupled multi-source system. A graphical comparison of these profiles, as shown in Fig.~\ref{EnergyTransferColor}, provides a visual reference for their spatial structure and serves as a practical guide for selecting appropriate transfer mechanisms in concrete models.

\section{Calculations}
\label{sec:calculations}

We implement the gravitational decoupling scheme. Our strategy is to adopt a well–behaved, isotropic Tolman~IV interior solution as the seed source \(T_{\mu\nu}\), and then use it to drive the geometric deformations \(\{f(r),g(r)\}\) that encode the coupling to a second gravitational source \(\theta_{\mu\nu}\). Once the deformation sector is obtained, we reconstruct the full two–fluid configuration and evaluate the resulting effective density and anisotropic pressures. Finally, we present radial profiles comparing: the seed (Tolman IV) quantities, the deformation–induced contributions, the combined effective variables \(\{\tilde{\rho},\tilde{p}_r,\tilde{p}_t\}\).

The line element of the seed configuration is written in the form
\begin{equation}
ds^2 = e^{\xi(r)} dt^2 - e^{\mu(r)} dr^2 - r^2 d\Omega^2,
\label{eq:tolmanIV_metric}
\end{equation}
with
\begin{equation}
e^{\xi(r)} =  B^2 \left(1 + \frac{r^2}{A^2} \right),
\label{eq:tolmanIV_xi}
\end{equation}
and
\begin{equation}
e^{-\mu(r)} = 
\frac{\left(1 - \dfrac{r^2}{C^2}\right)\left(1 + \dfrac{r^2}{A^2}\right)}
     {1 + \dfrac{2 r^2}{A^2}}.
\label{eq:tolmanIV_emu}
\end{equation}
Here \(A\), \(B\), and \(C\) are constants determined by matching to the exterior Schwarzschild configuration at the stellar boundary \(r=R\). No numerical specification of these constants is required at this stage.

For the Tolman~IV seed, the matter distribution is isotropic: \(p_r = p_t \equiv p\). The associated thermodynamic functions that solve the Einstein equations for the metric \eqref{eq:tolmanIV_metric}–\eqref{eq:tolmanIV_emu} are
\begin{equation}
\rho(r) = 
\frac{ 3 A^4 + A^2 \big( 3 C^2 + 7 r^2 \big) + 2 r^2 \big( C^2 + 3 r^2 \big) }
     { \kappa\, C^2 \big( A^2 + 2 r^2 \big)^2 },
\label{eq:tolmanIV_rho}
\end{equation}
\begin{equation}
p(r) = 
\frac{ C^2 - A^2 - 3 r^2 }
     { \kappa\, C^2 \big( A^2 + 2 r^2 \big) }.
\label{eq:tolmanIV_p}
\end{equation}
These expressions furnish the seed source \(T_{\mu\nu} = \mathrm{diag}[\rho,-p,-p,-p]\) used in the decoupling procedure.

Given \(T_{\mu\nu}\) as above, we introduce the second source \(\theta_{\mu\nu}\) through the geometric deformations
\begin{align}
\xi(r) &\longrightarrow \nu(r) = \xi(r) + g(r), \\
e^{-\mu(r)} &\longrightarrow e^{-\lambda(r)} = e^{-\mu(r)} + f(r),
\end{align}
which were defined in Eqs.~\eqref{eq:temporal_deform} and \eqref{eq:radial_deform}. Substituting these into the decoupled field equations for the deformation sector (Eqs.~\eqref{eq:E_theta}–\eqref{eq:Pt_theta}) yields a system that can be solved once an energy-exchange prescription \(\Delta E(r)\) is specified. The function \(g'(r)\) is related to \(\Delta E(r)\) through Eq.~\eqref{eq:energy_transfer}, while \(f(r)\) follows from the radial and tangential deformation equations.

After determining \(f(r)\) and \(g(r)\), we obtain the deformation–induced quantities
\(\{E(r),\mathcal{P}_r(r),\mathcal{P}_t(r)\}\) via
Eqs.~\eqref{eq:E_theta}–\eqref{eq:Pt_theta} and \eqref{eq:Z1}–\eqref{eq:Z2}. The full effective matter sector entering the total metric is then
\begin{align}
\tilde{\rho}(r) &= \rho(r) + E(r), \\
\tilde{p}_r(r) &= p(r) + \mathcal{P}_r(r), \\
\tilde{p}_t(r) &= p(r) + \mathcal{P}_t(r),
\end{align}
and the resulting anisotropy is \(\Pi(r) = \tilde{p}_t(r) - \tilde{p}_r(r)\).

\begin{figure*}[ht]
\center{\includegraphics[width=0.3\linewidth]{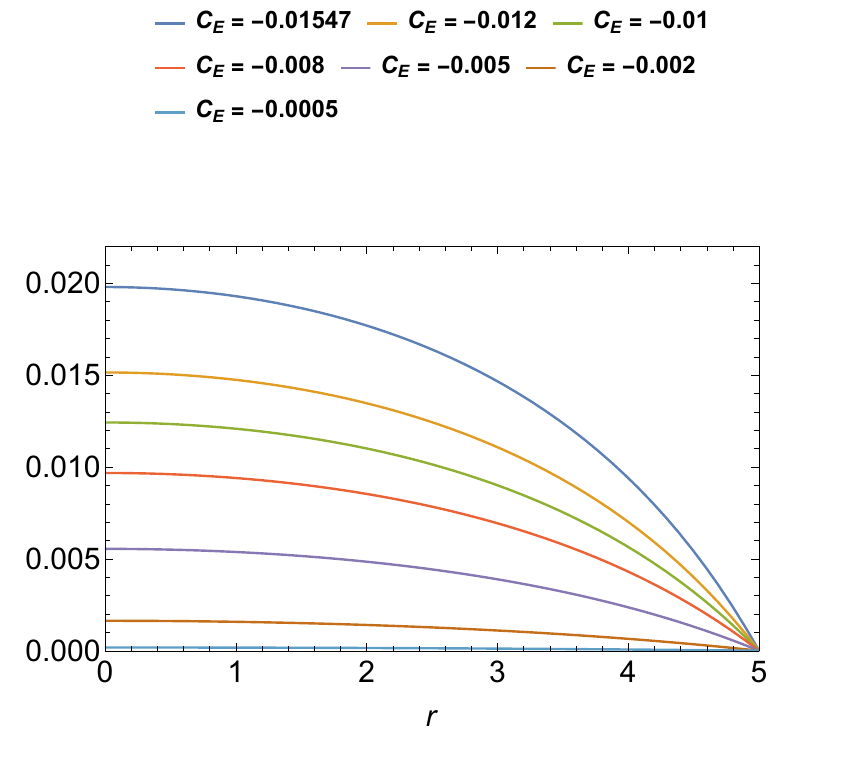}}\quad \includegraphics[width=0.3\linewidth]{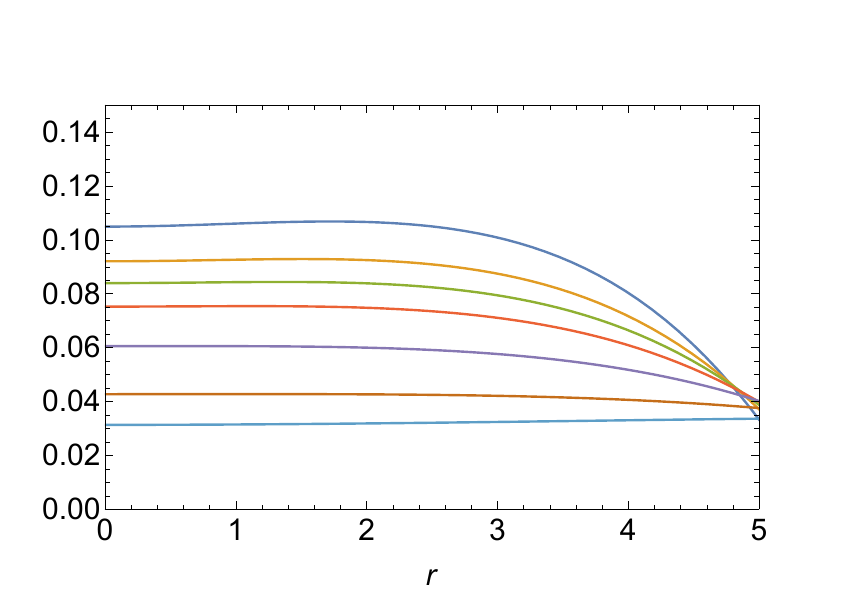}\quad
\includegraphics[width=0.3\linewidth]{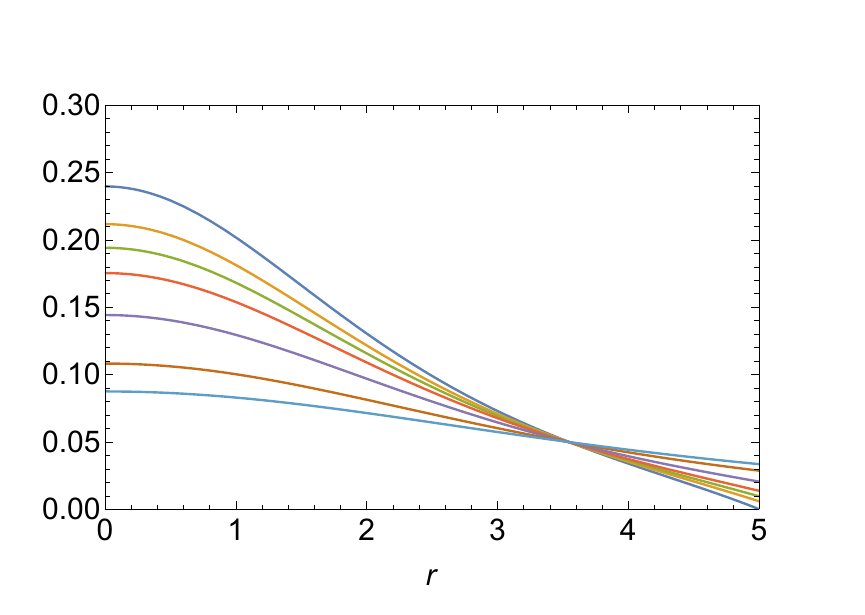}\\
\center{\includegraphics[width=0.3\linewidth]{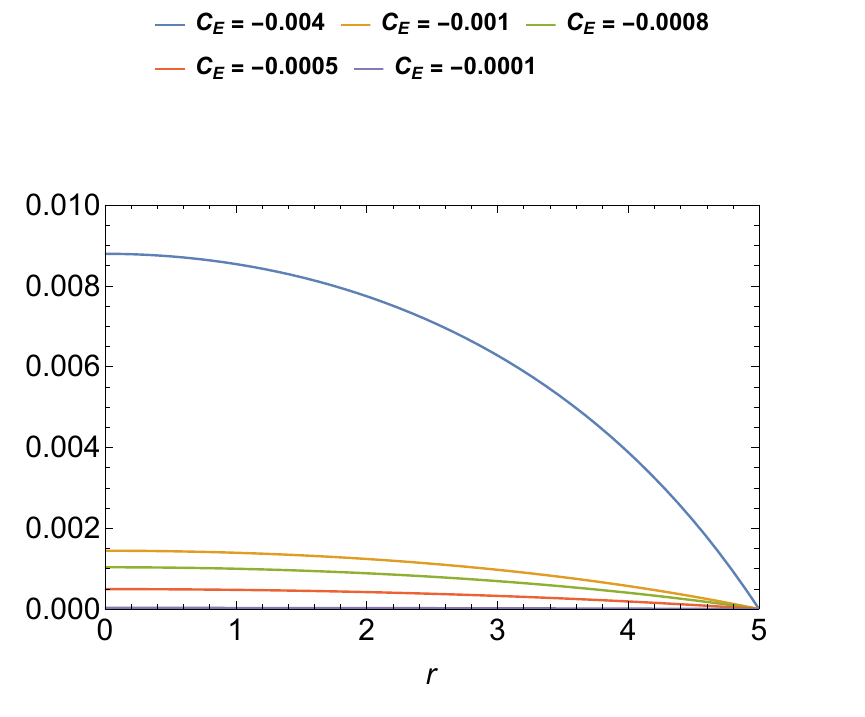}\quad \includegraphics[width=0.3\linewidth]{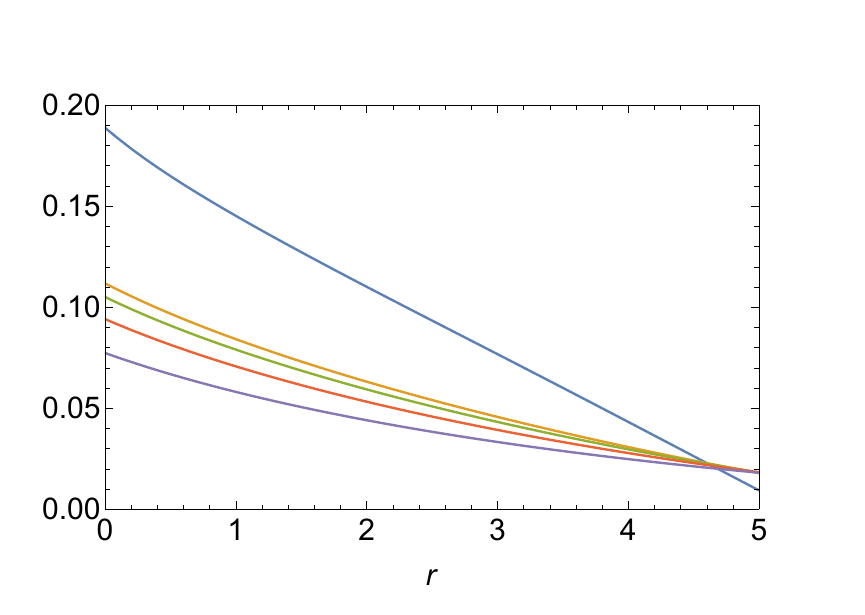}\quad \includegraphics[width=0.3\linewidth]{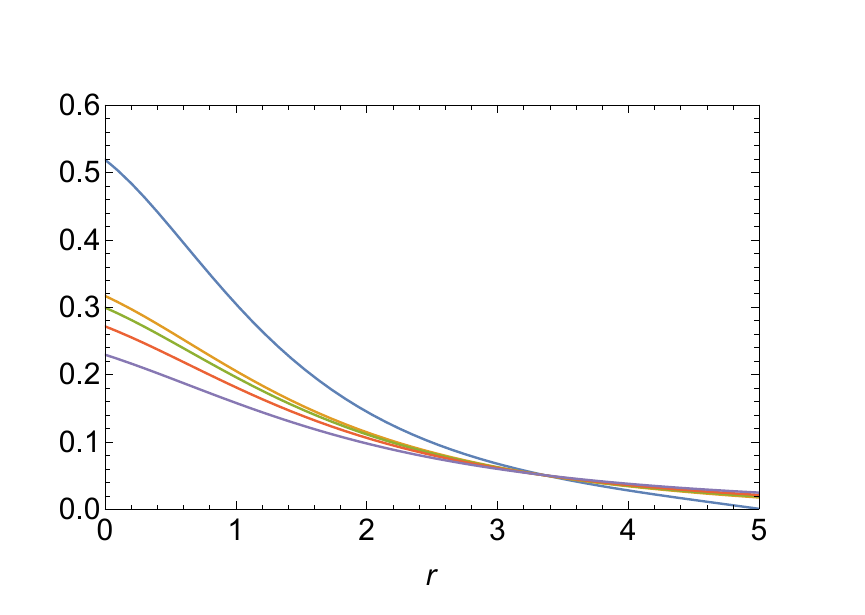}\\
\center{\includegraphics[width=0.3\linewidth]{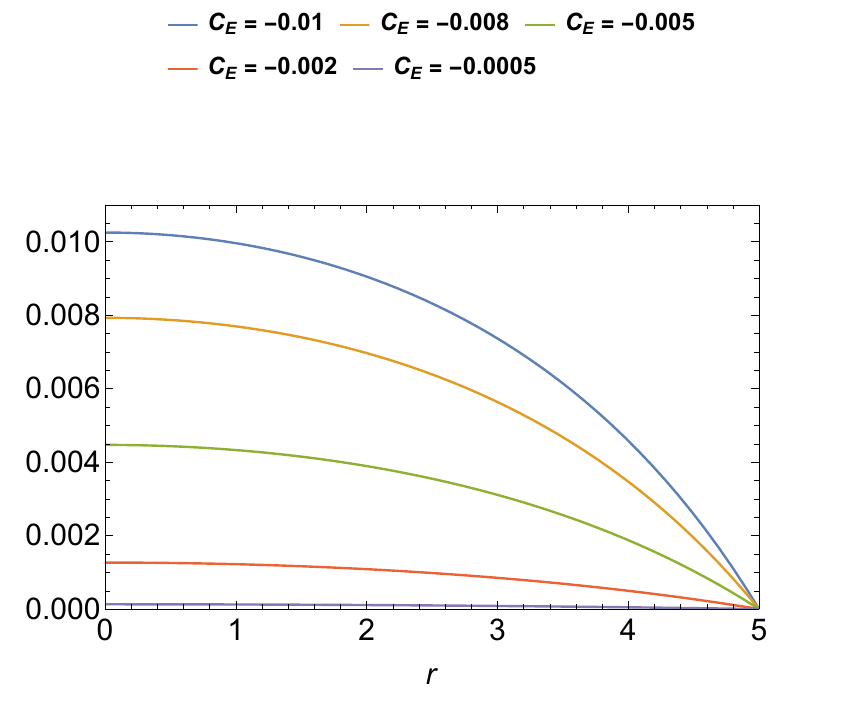}}\quad \includegraphics[width=0.3\linewidth]{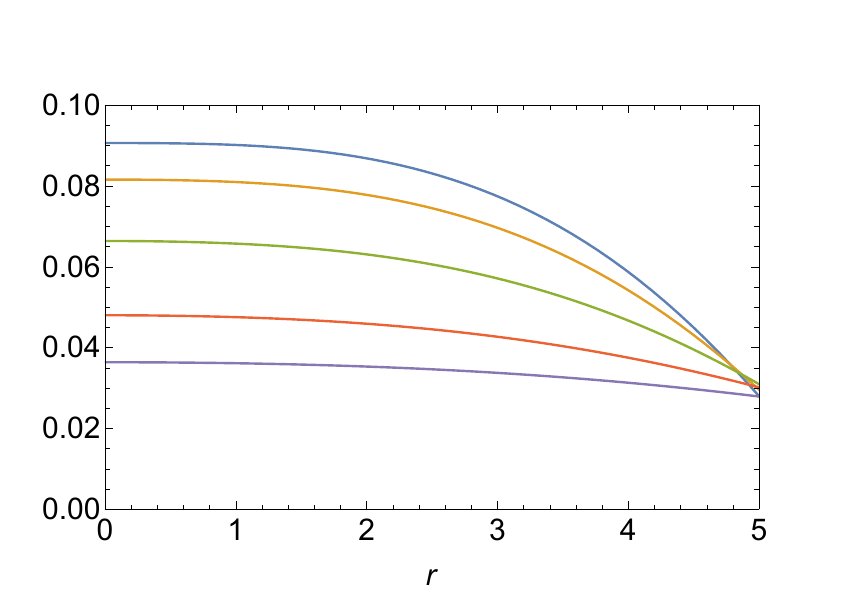}\quad \includegraphics[width=0.3\linewidth]{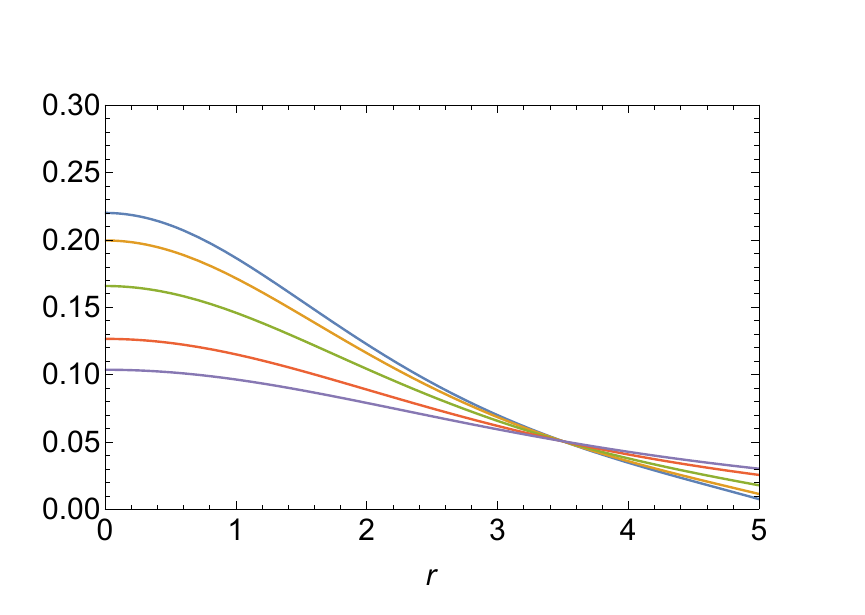}\\
\center{\includegraphics[width=0.3\linewidth]{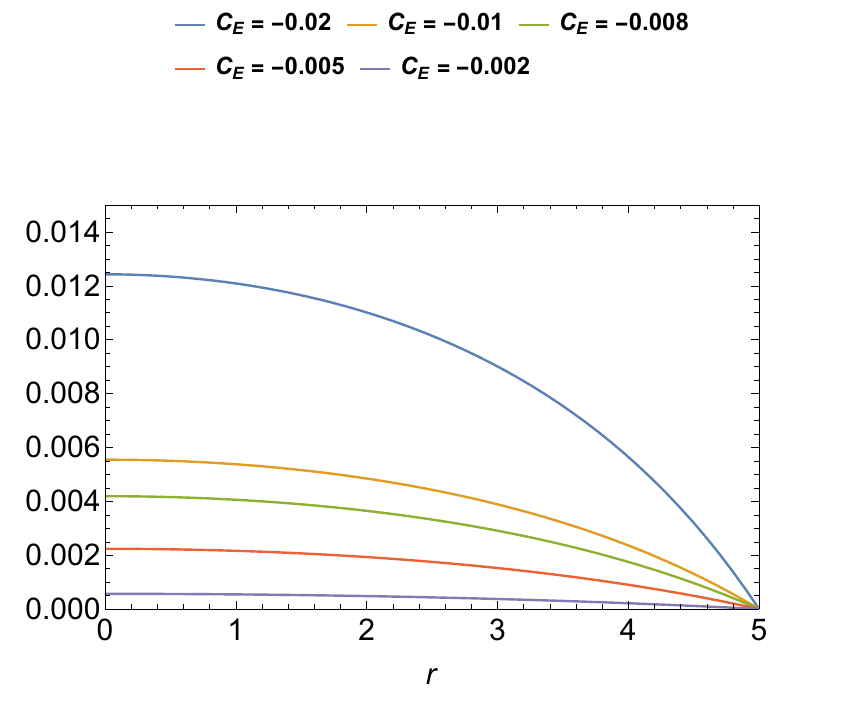}}\quad \includegraphics[width=0.3\linewidth]{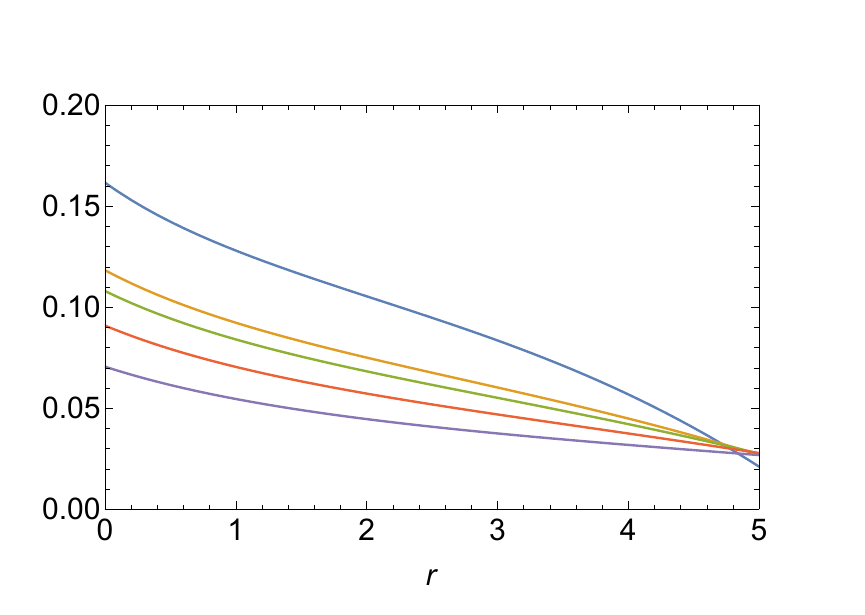}\quad \includegraphics[width=0.3\linewidth]{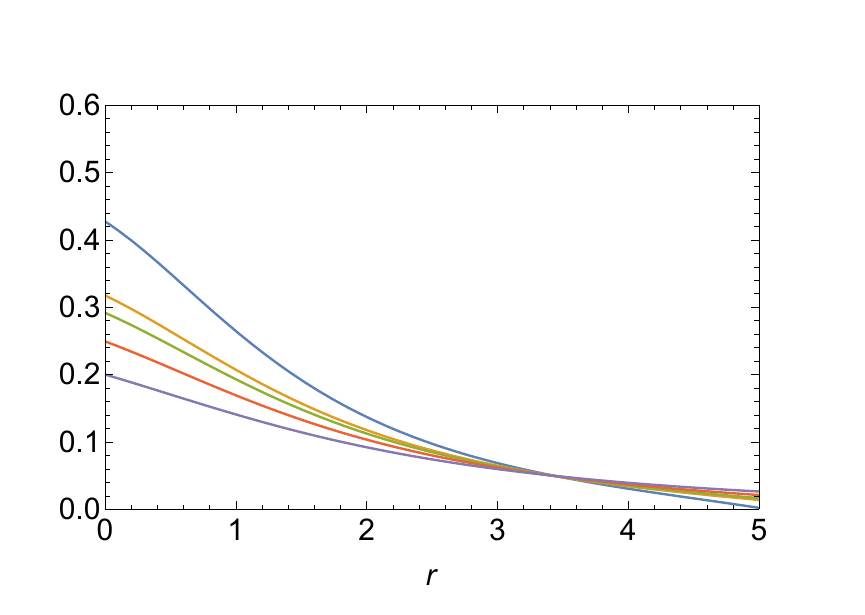}}
\caption{Radial pressure \(p_r(r)+\mathcal{P}_{r}(r)\) (left column), tangential pressure
\(p_t(r)+\mathcal{P}_{t}(r)\) (middle column), and energy density \(\rho(r)+E(r)\) (right
column) for different choices of the energy–transfer profile
\(\Delta E(r)\). From top to bottom, the rows correspond to the
power–law growth profile \(\Delta E_1\), the peaked profile
\(\Delta E_4\), the oscillatory profile \(\Delta E_7\), and the
saturating profile \(\Delta E_8\).}
\label{tolmanIV_profiles}
\end{figure*}

For each chosen energy-exchange ansatz \(\Delta E(r)\), we integrate the deformation equations using the Tolman~IV seed background \(\{\xi(r),\mu(r),\rho(r),p(r)\}\) given above. The integration constants are fixed by the matching conditions at \(r=R\), ensuring continuity with the exterior Schwarzschild solution and imposing vanishing total radial pressure at the surface. We then compute and plot the radial behavior of the seed, deformation, and effective total quantities for the energy density, radial pressure, and tangential pressure.

The exchange prescription \(\Delta E(r)\) introduced in
Eq.~\eqref{eq:energy_transfer} fixes the radial derivative of the temporal
deformation\,\(g(r)\):
\begin{equation}
g'(r)\;=\;\frac{2\,\Delta E(r)}{\rho(r)+p(r)},
\label{eq:gprime_def}
\end{equation}
where the seed quantities \(\rho(r)\) and \(p(r)\) are given in
Eqs.~\eqref{eq:tolmanIV_rho}--\eqref{eq:tolmanIV_p}.  Once \(g'(r)\) is
known, the radial geometric deformation \(f(r)\) follows directly from
the \((rr)\) component of the Einstein system (\(T\)-sector) combined
with the first deformation equation\,\eqref{eq:E_theta}:
\begin{equation}
f(r)\;=\;-\,e^{-\mu(r)}\;+\;
\frac{1}{1+r\!\left[\xi'(r)+g'(r)\right]}.
\label{eq:f_def}
\end{equation}
With the pair \(\{g'(r),f(r)\}\) fixed, every component of the
\(\theta_{\mu\nu}\)-sector is determined analytically.

From Eq.~\eqref{eq:E_theta} we obtain
\begin{equation}
E(r)\;=\;-\,
\frac{f(r)}{\kappa\, r^{2}}\;-\;
\frac{f'(r)}{\kappa\, r},
\label{eq:E_of_r}
\end{equation}
which contributes additively to the effective density
\(\tilde{\rho}(r)=\rho(r)+E(r)\).

The deformation‐induced pressures\footnote{  The symbols
\(\mathcal{P}_{r}\) and \(\mathcal{P}_{t}\) correspond to
\( \theta_{\,1}^{\;1}\) and \(\theta_{\,2}^{\;2}
      =\theta_{\,3}^{\;3}\), respectively.} are obtained from
Eqs.~\eqref{eq:Pr_theta}--\eqref{eq:Pt_theta}.  Using
Eqs.~\eqref{eq:gprime_def}--\eqref{eq:f_def},
\begin{align}
\mathcal{P}_{r}(r) &=
      \frac{C^{2}-A^{2}-3r^{2}}
           {\kappa\,C^{2}\bigl(A^{2}+2r^{2}\bigr)}
           \;-\;p(r),                                     \label{eq:Pr_of_r} \\[4pt]
\mathcal{P}_{t}(r) &= 
      \frac{e^{-\mu(r)}}{4\kappa}\!
      \Bigl[
          2g''+g'^{2}+\frac{2g'}{r}+2\xi'g'-\mu' g'
      \Bigr]                                         \notag\\
 &\quad +\,\frac{f(r)}{4\kappa}\!
      \Bigl[
          2\nu''+\nu'^{2}+\frac{2\nu'}{r}
      \Bigr]
      +\frac{f'(r)}{4}\Bigl[\nu'+\tfrac{2}{r}\Bigr].
\label{eq:Pt_of_r}
\end{align}
Here \(\nu(r)=\xi(r)+g(r)\) and primes denote
\(d/dr\).  The effective pressures are then
\(\tilde{p}_{r}=p+\mathcal{P}_{r}\) and
\(\tilde{p}_{t}=p+\mathcal{P}_{t}\).

Equations~\eqref{eq:gprime_def}--\eqref{eq:Pt_of_r}, together with the
matching conditions close the system:
\begin{enumerate}
\item the seed Tolman\,IV solution
      \(\{\xi,\mu,\rho,p\}\) satisfies the Einstein equations;
\item the deformations
      \(\{g',f\}\) follow from the chosen \(\Delta E(r)\);
\item the second source
      \(\{E,\mathcal{P}_{r},\mathcal{P}_{t}\}\)
      satisfies its field equations and guarantees the total
      conservation law \(\nabla_{\mu}\tilde{T}^{\mu\nu}=0\).
\end{enumerate}

\begin{figure*}[ht]
\center{
\includegraphics[width=0.23\linewidth]{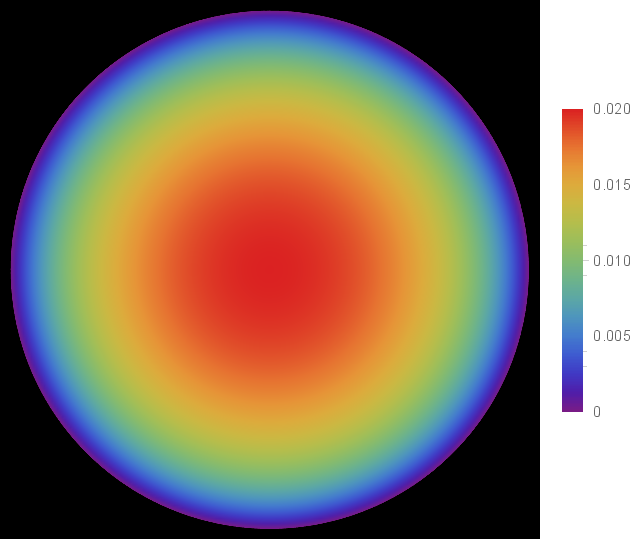}\quad
\includegraphics[width=0.23\linewidth]{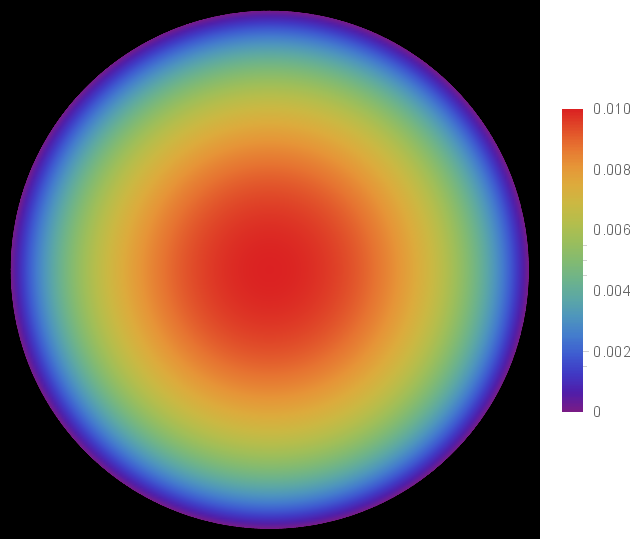}\quad 
\includegraphics[width=0.23\linewidth]{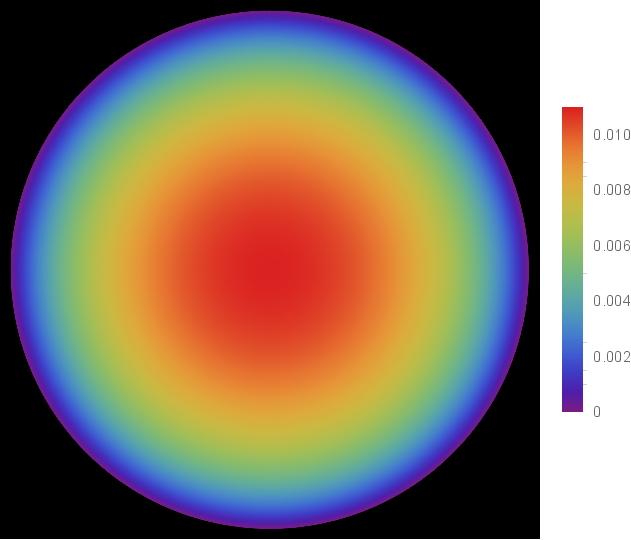}\quad 
\includegraphics[width=0.23\linewidth]{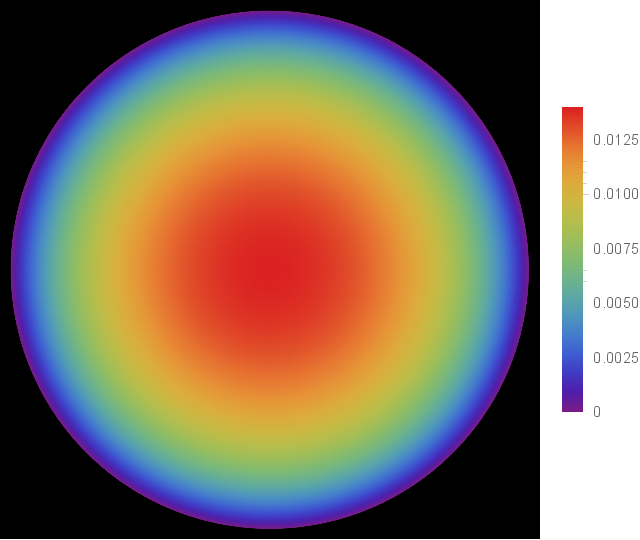}}\\
\center{
\includegraphics[width=0.23\linewidth]{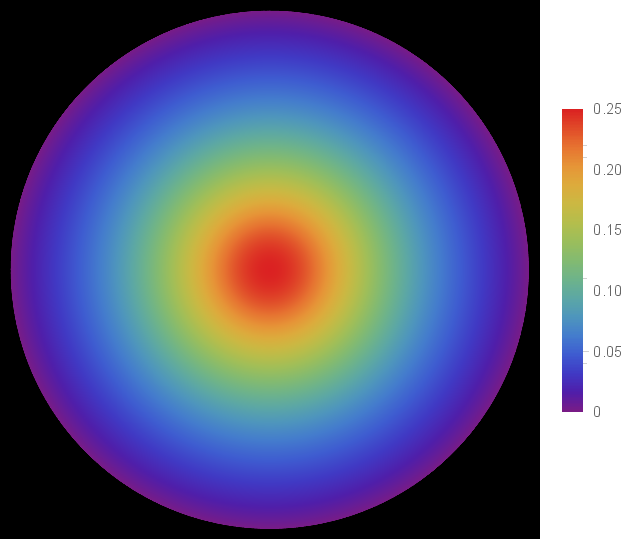}\quad
\includegraphics[width=0.23\linewidth]{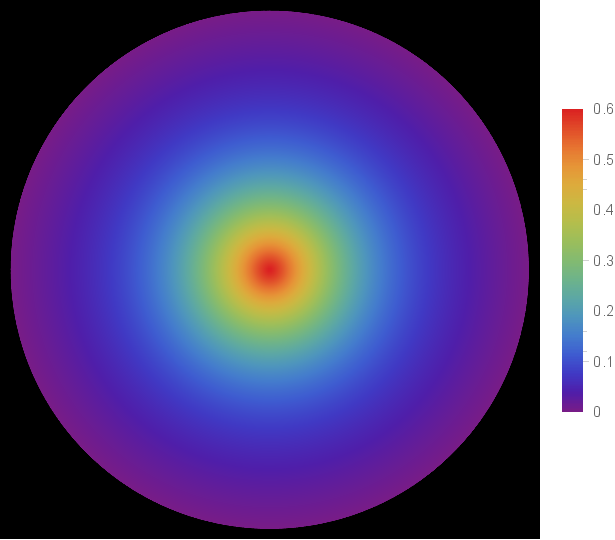}\quad
\includegraphics[width=0.23\linewidth]{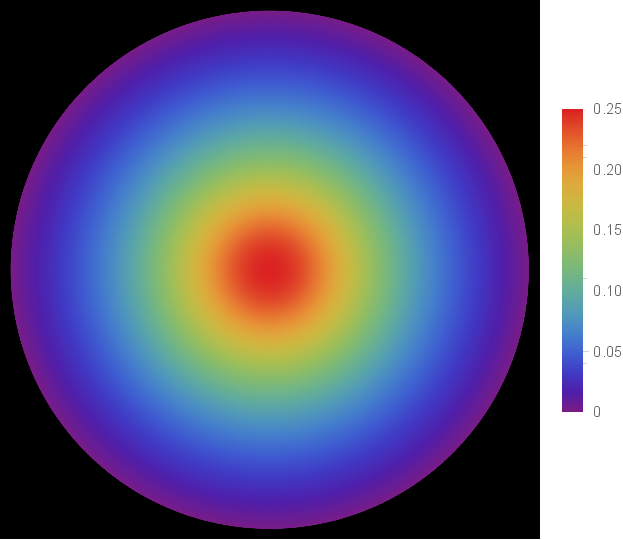}\quad
\includegraphics[width=0.23\linewidth]{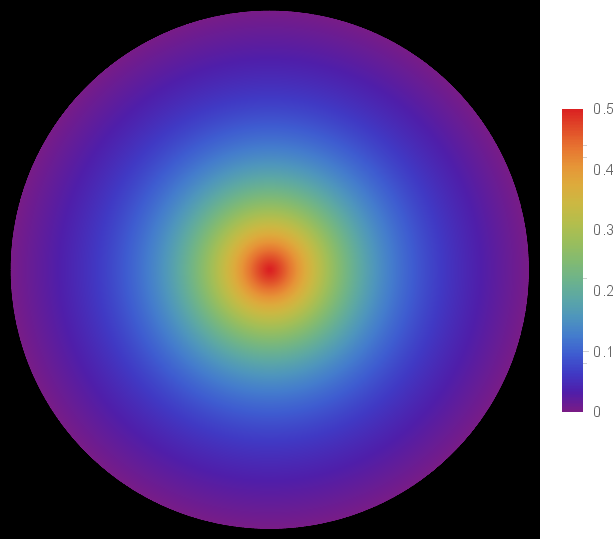}}\\
\caption{Radial pressure (top row) and energy density (bottom row) of the
highest curve of each selected energy transfer function. From left
to right, each column corresponds to the power–law growth profile
\(\Delta E_1\), the peaked profile \(\Delta E_4\), the oscillatory
profile \(\Delta E_7\), and the saturating (sigmoidal) profile
\(\Delta E_8\).}
\label{tolmanIV_profiles_Energy}
\end{figure*}

With all functions specified, we compute and plot the
combined profiles
\(\bigl\{\tilde{\rho}(r),\,\tilde{p}_{r}(r),\,\tilde{p}_{t}(r)\bigr\}\)
for each exchange model \(\Delta E(r)\).  The resulting curves,
displayed in Fig.~\ref{tolmanIV_profiles} and ~\ref{tolmanIV_profiles_Energy} reveal how different
energy‑transfer mechanisms reshape the interior density and anisotropic
pressure structure of the star. Only four energy transfer functions give regular solutions.

\section{Conclusion}
\label{sec:conclusion}

In this work we have extended the Gravitational Decoupling programme to the case in which one of the interacting constituents obeys a polytropic equation of state while the companion source is left unspecified.
Our main goal was to understand, in a fully analytic setting, how distinct prescriptions for the local exchange of energy, $\Delta E(r)$, reshape the interior structure of a static, spherically–symmetric self‑gravitating configuration.

\begin{enumerate}
\item Starting from a well–behaved Tolman IV seed solution, we derived closed‑form expressions for the geometric deformations $\{f(r), g(r)\}$ that encode the backreaction of an auxiliary source $\theta_{\mu\nu}$.
The temporal deformation is set directly by $\Delta E(r)$ through the compact relation $g' = \frac{2 \, \Delta E}{\rho + p_r}$, while the radial deformation follows algebraically from the Einstein equations.

\item Eight physically motivated ansätze for the energy‑transfer function were analysed (power‑law, inverted power‑law, exponential, peaked, logarithmic, smooth decay, oscillatory and sigmoidal).
Regularity of the effective variables, fulfillment of the matching conditions at the stellar surface, and absence of central singularities were used as acceptance criteria.
Only four profiles—$\Delta E_2$ (inverted power‑law), $\Delta E_3$ (exponential), $\Delta E_6$ (smooth decay), and $\Delta E_8$ (sigmoidal)—satisfied the requirements for a regular solution.

\item For each admissible profile we obtained analytic formulas for the effective density $\tilde{\rho}(r)$, the radial and tangential pressures $\tilde{p}_r(r)$ and $\tilde{p}_t(r)$, and the anisotropy $\Pi(r)$.
Representative radial behaviours were displayed in Fig.~\ref{tolmanIV_profiles}, illustrating how the location and intensity of the exchange layer correlate with: an increase or decrease of the central density, the onset of pressure anisotropy—vanishing in the seed model but revived by GD, and the shifting of the vanishing‑pressure radius that defines the stellar boundary.

\item The Israel–Darmois junction conditions were enforced exactly, yielding analytic relationships between the model parameters $\{A, B, C, C_E, n\}$ and the global mass–radius pair $(M, R)$.
Those relations guarantee that our composite interior matches smoothly to the Schwarzschild exterior and remains free of thin shells or surface layers.
\end{enumerate}

The analysis shows that a polytropic sector can significantly modify the macroscopic properties of a relativistic star when it channels energy to—or draws energy from—an additional fluid.
Depending on the shape of $\Delta E(r)$, the total mass can increase or decrease with respect to the pure Tolman IV configuration, and the stellar radius can either expand or contract.
Moreover, the induced anisotropy can reach magnitudes comparable to those discussed in models of pressure‑supported quark stars or magnetised compact objects.
This highlights the importance of treating energy–momentum exchange explicitly rather than subsuming it into an effective single‑fluid description.

Although all solutions presented here satisfy the standard energy and causality conditions, a detailed stability analysis—e.g.\ via the Harrison–Zeldovich–Novikov criterion or the study of radial perturbations—lies beyond the present scope.
Preliminary inspection indicates that the smooth‑decay and sigmoidal profiles offer the most promising ground for stable equilibria, as they avoid steep gradients in the effective sound speed.
We defer a systematic treatment of dynamical stability to future work.

The framework developed here can be generalised in several directions:
\begin{itemize}
\item inclusion of electric charge or magnetic fields in one (or both) sectors, allowing the study of charged anisotropic stars with explicit energy exchange;
\item extension to slowly rotating configurations by implementing the Hartle–Thorne formalism on top of GD, which would link internal transfer mechanisms to observable moments of inertia;
\item application to alternative theories of gravity by letting $\theta_{\mu\nu}$ encode, for instance, effective Einstein–Gauss–Bonnet corrections;
\item exploration of time‑dependent exchange functions $\Delta E(t, r)$ to model transient phenomena such as phase transitions or thermal relaxation in proto‑neutron stars.
\end{itemize}

In summary, gravitational decoupling augmented by an explicit, radial‑dependent energy‑transfer function offers a powerful analytic avenue to probe the interplay between multiple fluids inside compact objects.
Our results demonstrate that even moderate exchanges can leave clear imprints on the density and pressure profiles, potentially discernible through precise astrophysical measurements of mass, radius and tidal deformability.
We anticipate that the techniques presented here will facilitate realistic multi‑component modelling in both general relativity and its extensions, bridging the gap between idealised single‑fluid solutions and the rich microphysics of high‑density matter.

\begin{acknowledgments}

The author would like to acknowledge the Research
Center for Theoretical Physics and Astrophysics and Institute of Physics of Silesian University in Opava for institutional support. 

\end{acknowledgments}

\bibliography{references}

\end{document}